\documentclass[fleqn,11pt]{article}
\long\def\symbolfootnote[#1]#2{\begingroup%
	\def\thefootnote{\fnsymbol{footnote}}\footnote[#1]{#2}\endgroup} 
\usepackage{mathtools}
\usepackage{amsmath}
\usepackage{amssymb,enumerate,color}
\usepackage{dsfont}
\usepackage{blindtext}
\usepackage{booktabs}
\usepackage{bm}
\usepackage{nccmath}
\usepackage{apacite}
\usepackage{rotating}

\usepackage{caption}
\usepackage{subcaption}

\usepackage{grffile}

\usepackage{graphicx}

\usepackage{lscape}

\usepackage{blindtext}

\usepackage{amssymb}
\usepackage{varioref} 
\usepackage{textcomp}

\usepackage{booktabs}

\usepackage{multirow}

\usepackage{pdflscape}

\usepackage{relsize}
\usepackage{tablefootnote}
\usepackage{wrapfig}
\usepackage{natbib}
\usepackage{listings}
\usepackage{longtable}
\usepackage{url}
\usepackage{tabularx}
\usepackage{float}
\usepackage{epsfig}
\usepackage[titletoc]{appendix}
\usepackage{arydshln}
\usepackage{footnote}
\usepackage{secdot}
\usepackage{bigints}
\usepackage[utf8]{inputenc}
\usepackage[english]{babel}
\usepackage{xr}
\usepackage{adjustbox}
\usepackage{graphicx} 

\usepackage{multicol}
\usepackage{enumitem}

\usepackage{bigstrut}
\usepackage{rotating}
\usepackage{xr}

\usepackage{float}

\newcommand{\sumk}{\sum_{k=0}^{T}}

\title{\bf{A New Cure Rate Model with Discrete and Multiple Exposures}}

\author{\bf{Suvra Pal\footnote{Corresponding author. E-mail address: suvra.pal@uta.edu Tel: (817) 272-7163}}\\[1ex]
	Department of Mathematics, University of Texas at Arlington, Arlington, \\
	Texas 76019, USA}
\date{}

\begin{document}
	\maketitle
	
\begin{abstract}

\noindent Cure rate models are mostly used to study data arising from cancer clinical trials. Its use in the context of infectious diseases has not been explored well. In 2008, Tournoud and Ecochard first proposed a mechanistic formulation of cure rate model in the context of infectious diseases with multiple exposures to infection. However, they assumed a simple Poisson distribution to capture the unobserved pathogens at each exposure time. In this paper, we propose a new cure rate model to study infectious diseases with discrete multiple exposures to infection. Our formulation captures both over-dispersion and under-dispersion with respect to the count on pathogens at each time of exposure. We also propose a new estimation method based on the expectation maximization algorithm to calculate the maximum likelihood estimates of the model parameters. We carry out a detailed Monte Carlo simulation study to demonstrate the performance of the proposed model and estimation algorithm. The flexibility of our proposed model also allows us to carry out a model discrimination. For this purpose, we use both likelihood ratio test and information-based criteria. Finally, we illustrate our proposed model using a recently collected data on COVID-19.\\
	

\end{abstract}
	
\section{Introduction}

The study of models for survival or time-to-event data with a surviving fraction, known as cure rate models, have been extensively studied during the past few decades. Cure rate models have become more prevalent as new treatments for diseases cause a larger population of patients who are permanently cured of their disease. The first cure rate model was proposed by \citet{Boa49} and later revised by \citet{Ber52}. They proposed a two-component mixture model to analyze data where some subjects in the study were not susceptible to the event of interest. These subjects are called immunes, cured, or long-term survivors. \citet{Yak93} created a new formulation for mixture models to more effectively study cancer metastases. \citet{Tso98} further advanced the study of mixture models with the development of bounded cumulative hazard models. Other recent works of note include \citet{Davies21}, \citet{Pal21SIM}, \citet{Mil22}, \citet{pal2022stochastic}, \citet{Wang22}, \citet{Paletal23}, \citet{PalSVM23} and \citet{Ase23}, amongst others. Interested readers may also look at \citet{PengYu21} for a book-length account on mixture cure models. However, in the works previously mentioned, the models described primarily dealt with cancer metastasis and the spread of tumors in patients. Until recently, the application of these models in the context of infectious disease modeling had not been well researched. Some of the first work looking into the application of these models in the context of infectious diseases was done by \citet{Tou07}, who studied a cure model with multiple discrete infection occasions. Specifically, the authors proposed a cure model where the number of pathogens affecting the $i^{th}$ patient, $i=1,2,...,n$, at exposure time $t_k$, $k=0,1,2,...,T$ ($k=0$ representing the time to initial exposure), was modeled using a Poisson random variable with mean $\theta_{t_k}$. This allowed them to study mother to child transmission of HIV-1 virus and the spread of nosocomial infections. Although the Poisson distribution was used to capture the latent number of pathogens at exposure time $t_k$, there was no justification provided for the suitability of the Poisson distribution, given that it cannot handle over- and under-dispersion that we commonly encounter with respect to the count in the number of pathogens. To advance their work, \citet{Tou08} again proposed a cure model with multiple infection occasions, but considered different distributions to model the number of pathogens affecting the $i^{th}$ patient at time $t_k$. They considered modeling the number of pathogens using a compound Poisson distribution to consider inter-subject heterogeneity. The issue with the compound Poisson distribution is the difficulty in interpreting a distribution in which a discrete random variable is attained using the sum of continuous random variables. In order to overcome these drawbacks, we propose the use of the Conway-Maxwell Poisson (COM-Poisson) distribution to model the number of pathogens at each infection occasion $t_k$, $k=0,1,2,...,T$, assuming multiple infection occasions. This results in a flexible model to study infectious diseases with multiple discrete exposures and is more realistic compared to the model studied by \citet{Tou07} and \citet{Tou08}. Furthermore, if we consider the special case of just one exposure, i.e. $k=0$, our model reduces to the COM-Poisson cure rate model of \citet{Rod09} and \citet{Bal16}. To develop the likelihood inference, we do not rely on the numerical maximization of the log-likelihood function. Instead, motivated by the introduction of missing data arising from the set of censored observations, we develop the steps of the expectation maximization (EM) algorithm to calculate the maximum likelihood estimates (MLEs) of the model parameters.  

The COM-Poisson distribution was first discussed in \citet{Con62} to help solve problem related to queuing systems. Since its introduction, many authors have studied its properties including \citet{Shm05}, \citet{Boat03}, and more recently \citet{Li20b}. The COM-Poisson is a generalized discrete distribution that allows for under- and over- dispersion of count data. As such, the COM-Poisson distribution allows for far more flexibility in modeling count data than other commonly used distributions. The Poisson, geometric, and Bernoulli distributions are all special cases of the COM-Poisson distribution. \citet{Rod09} were one of the first to explore the uses of the COM-Poisson distribution in the context of cure rate model. In the work of \citet{Rod09}, the authors used the COM-Poisson distribution as a way to form a more flexible version of the unified cure rate model that was originally formulated by \citet{Yin05}. For other recent works on the COM-Poisson distribution in the context of cure rate models together with flexible modeling of lifetime data, interested readers may look at \citet{Bal13,Bal15b}, \citet{Pal17c}, \citet{Pal18}, \citet{Wia18}, and \citet{Majak19}, among others.

The rest of this paper is organized as follows. In Section 2, we introduce the COM-Poisson cure rate model in the context of infectious diseases with discrete multiple exposures to infection. In Section 3, we develop the estimation algorithm based on the EM algorithm to calculate the MLEs of the model parameters. In Section 4, we carry out an extensive Monte Carlo simulation study, where we present the results of both model fitting and model discrimination. To carry out the model discrimination, we use both likelihood ratio test and information-based criteria. In Section 5, we illustrate our proposed model using a recently collected data on COVID-19. Finally, in Section 6, we make some concluding remarks and discuss some potential future research work in the context of infectious diseases.

\section{COM-Poisson cure rate model with discrete multiple exposures}
	
Let $n$ represent the number of patients in the study. Let $t=\lbrace t_0,...,t_k,...,t_T\rbrace$ denote the multiple and successive moments of infections with $t_0$ denoting the initial moment of infection. At each moment of infection, let $M_{i,t_k}$, for $1\leq i\leq n$, $t_0\leq t_k\leq t_T$, denote the number of pathogens infecting the $i^{th}$ patient at time $t_k$. Note that $M_{i,t_k}$ are unobservable variables and we assume it to follow a discrete distribution with probability mass function (pmf) $p_{m_{i,t_k}}$. We assume the number of pathogens at each exposure time to be independent, which is along the lines of \citet{Tou08}. It is important to note that we do not assume the distributions of $M_{i,t_0}, M_{i,t_1},..., M_{i,t_T}$ to be identical for the $i^{th}$ patient. Rather, at each exposure time $t_k$, we assume a suitable parameter of the distribution to depend on the exposure time. Let $Z_{i,j,t_k}$, for $1\leq i\leq n$, $t_0\leq t_k\leq t_T$ and $0\leq j\leq M_{i,t_k}$, be the $j^{th}$ pathogen promotion time of the $i^{th}$ patient at time $t_k$, in other words, the time taken by the $j^{th}$ pathogen to produce the event, which in the context of infectious diseases may represent the first biological sign of infection. For a given $M_{i,t_k}$, we assume $Z_{i,j,t_k}$ to be distributed with cumulative distribution function (cdf) $F(z)=F(z|\bm{\gamma})=1-S(z|\bm{\gamma}),$ where $\bm{\gamma}$ is the associated vector of parameters and $S(\cdot)$ is the corresponding survival function. Since this model follows data with multiple exposures, we need to re-parameterize the time scale of infection on the original scale. Let $y=z+t_k$, then, we can re-parameterize the cdf and survival function as follows:
                 \begin{eqnarray}
			F_{t_k}(y|\bm{\gamma})= \begin{cases}
			F(y-t_k), & y-t_k>0 \\
			0, & y-t_k\leq 0 \\
			\end{cases}
			\label{F}
		\end{eqnarray}	
	and
	\begin{eqnarray}
			S_{t_k}(y|\bm{\gamma})= \begin{cases}
			S(y-t_k), & y-t_k>0 \\
			1, & y-t_k\leq 0. \\
			\end{cases}
		\label{S}
	\end{eqnarray}
	\indent Given that patients are exposed at different discrete time points and at each exposure time there are several competing pathogens, the time-to-event, which can be the first biological sign of infection or time to recovery from an infectious disease, can be defined as
	\begin{equation}
			Y_i=\text{min}\lbrace Z_{i,j,t_k}, 0\leq j\leq M_{i,t_k}, t_0\leq t_k\leq t_T\rbrace,\ \  i=1,2,...,n,
		\label{Y}
	\end{equation}
	where $Z_{i,0,t_k}$ is such that
	$$P(Z_{i,0,t_k}=\infty)=1,\hspace{5mm} t_0\leq t_k\leq t_T.$$ 
	A patient is termed “immune” of an infection if there are no competing pathogens at each exposure time point, and its probability, termed as cure rate, is defined as
	$$p_{0i}=P(M_{i,t_k}=0,\forall t_0\leq t_k\leq t_T),\hspace{3mm}i=1,2,...,n.$$
	\indent We will now introduce the COM-Poisson distribution as a means of modeling the number of competing pathogens at each discrete exposure time point. For this purpose, we will first introduce the COM-Poisson distribution. If a random variable $M$ follow a COM-Poisson distribution, then, its pmf is given by
	$$P[M=m;\theta,\nu]=\frac{1}{Z(\theta,\nu)}\frac{\theta^m}{(m!)^\nu},\hspace{5mm}m=0,1,2,...,$$
	where $\theta>0$, $\nu>0$, and  $Z(\theta,\nu)$ is the normalizing constant calculated by
	$$Z(\theta,\nu)=\sum_{j=0}^{\infty}\frac{\theta^j}{(j!)^\nu}.$$
	From the above equation, we can see the cured fraction, in other words the part of the population that is not affected by the pathogen, denoted by $p_0$, is given by
	$$p_0=P[M=0;\theta,\nu]=\frac{1}{Z(\theta,\nu)}.$$
The COM-Poisson distribution has three special cases. When $\nu=1$, then, $Z(\theta,\nu)=e^{\theta}$, which results in the Poisson distribution with mean $\theta$. As $\nu\rightarrow\infty$, $Z(\theta,\nu)\rightarrow1+\theta$, which means the COM-Poisson distribution converges in distribution to the Bernoulli distribution with $P[M=1;\theta,v]=\frac{\theta}{1+\theta}$. When $\nu=0$ and $\theta<1$, $Z(\theta,\nu)=\frac{1}{1-\theta}$, and the COM-Poisson reduces to a geometric distribution with parameter $1-\theta$. However, if $\nu=0$ and $\theta\geq1$, then $Z(\theta,\nu)$ does not converge. Therefore, the COM-Poisson distribution is undefined in this particular case. One advantage of the COM-Poisson distribution is that it allows for both under-dispersion and over-dispersion of data relative to the Poisson distribution. When $\theta>1$, the data is under-dispersed relative to the Poisson distribution, whereas when $\theta<1$, the data is over-dispersed relative to the Poisson distribution.
	
In this research, we will assume the number of competing pathogens to follow a COM-Poisson distribution with parameters $\theta_{t_k}$ and $\nu$ at exposure time $t_k$, for $k=0,1,2,…,T$. Note that the parameter $\theta_{t_k}$ represents the infection intensity at exposure time $t_k$ and hence carries biological interpretation. In a practical scenario, the infection intensity at a given exposure time will differ across patients. To capture this heterogeneity in patient population, we link $\theta_{t_k}$ to a set of covariates $\boldsymbol x_{t_k}$ at each exposure time $t_k$, for $k=0,1,2,…,T$, using the log-linear link function $\theta_{t_k} = \exp(\boldsymbol x_{t_k}^\prime \boldsymbol\beta_{t_k})$, where $\boldsymbol\beta_{t_k}$ is the corresponding vector of regression coefficients.
Now, let us consider two exposure times, $t_0$ and $t_1$, and derive the survival function of the random variable $Y$ in \eqref{Y}, known as the population survival function or the long-term survival function.

\newtheorem{prop}[theorem]{Theorem}
\begin{prop}
Given two discrete exposure times $t_0$ and $t_1$, let $M_{t_0}$ and $M_{t_1}$ denote the number of pathogens at times $t_0$ and $t_1$. Let $M_{t_0}$ and $M_{t_1}$ both follow COM-Poisson distribution with parameters $(\theta_{t_0},\nu)$ and $(\theta_{t_1},\nu)$, respectively. Furthermore, let $S_{t_0}(y)$ and $S_{t_1}(y)$ denote the pathogen promotion time distribution at exposure times $t_0$ and $t_1$, respectively, as described in \eqref{S}. Then, the overall survival function of the variable $Y$ in \eqref{Y}, denoted by $S_{pop}(y)$, is given by
\begin{equation}
S_{pop}(y)=P[Y > y]=\frac{Z(\theta_{t_0}S_{t_0}(y),\nu)}{Z(\theta_{t_0},\nu)}\frac{Z(\theta_{t_1}S_{t_1}(y),\nu)}{Z(\theta_{t_0},\nu)}.
\end{equation}
\label{th1}
\end{prop}
A proof of Theorem \ref{th1} is provided in the Appendix.\\
	
If we generalize the above survival function with multiple exposure times $\lbrace t_0,…,t_k,…,t_T\rbrace$, then, we have
	\begin{eqnarray}
			S_{pop}(y)=\prod_{k=0}^{T}\frac{Z(\theta_{t_k}S_{t_k}(y),\nu)}{Z(\theta_{t_k},\nu)}.
			\label{Sp}
	\end{eqnarray}
	Note that in \eqref{Sp}, if we just consider one exposure, then, $S_{pop}(y) = \frac{Z(\theta_{t_0}S_{t_0}(y),\nu)}{Z(\theta_{t_0},\nu)}$, which reduces to the model proposed by \citet{Rod09} and \citet{Bal16}. The density function corresponding to \eqref{Sp}, known as the long-term density function, is given by
	\begin{equation}
			f_{pop}(y)=-S'_{pop}(y)=\sum_{k=0}^{T}\bigg[\frac{1}{Z(\theta_{t_k},\nu)}\frac{f_{t_k}(y)}{S_{t_k}(y)}\sum_{j=1}^{\infty}\frac{j\lbrace \theta_{t_k} S_{t_k}(y)\rbrace^j}{(j!)^v}\prod_{\stackrel{i\neq k}{i=0}}^{T}\frac{Z(\theta_{t_i}S_{t_i}(y),\nu)}{Z(\theta_{t_i},\nu)}\bigg],
	\label{fp}
	\end{equation}
	where $f_{t_k}(y)$ is the density function associated with $S_{t_k}(y)$. Hence, the cure rate or cured proportion is given by
	\begin{equation}
			p_0=S_{pop}(\infty)=P[M_{t_k}=0\hspace{3mm} \forall k\in\lbrace0,1,2,...,T\rbrace]=\prod_{k=0}^{T}\frac{1}{Z(\theta_{t_k},\nu)}.
	\label{p0}
	\end{equation}
	In Table \ref{table:T1}, we present the long-term survival function and the cured proportion for the COM-Poisson cure rate model and its special cases with discrete multiple exposures. In Table \ref{table:T2}, we present the corresponding long-term density functions.
		\begin{table}[H]
			\centering
			\caption{Expressions of long-term survival function and cured portion for the COM-Poisson cure rate model and its special cases with discrete multiple exposures}
			\begin{tabular}{|| c c c ||}
				\hline
				Model    &  $S_{pop}(y)$ & $p_0$ \\
				\hline
				COM-Poisson     & $\prod_{k=0}^{T}\frac{Z(\theta_{t_k}S_{t_k}(y),\nu)}{Z(\theta_{t_k},\nu)}$    & $\prod_{k=0}^{T}\frac{1}{Z(\theta_{t_k},\nu)}$     \\[1ex]
				Poisson & $\text{exp}\Big[\sum_{k=0}^{T}\lbrace-\theta_{t_k}F_{t_k}(y)\rbrace\Big]$ & $\text{exp}\Big[\sum_{k=0}^{T}\lbrace-\theta_{t_k}\rbrace\Big]$          \\[1ex]
				Bernoulli    & $\prod_{k=0}^{T} \Big\{ \frac{1+\theta_{t_k}S_{t_k}(y)}{1+\theta_{t_k}}   \Big\}$  & $\prod_{k=0}^{T} \big\{\frac{1}{1+\theta_{t_k}}  \big\}$    \\[1ex]
				Geometric & $\prod_{k=0}^{T}\frac{1-\theta_{t_k}}{1-\theta_{t_k}S_{t_k}(y)}$ & $\prod_{k=0}^{T}(1-\theta_{t_k})$\\[1ex]
				\hline
			\end{tabular}
		\label{table:T1}
		\end{table}
	
		\begin{table}[H]
			\centering	
			\caption{Expressions of long-term density function for the COM-Poisson cure rate model and its special cases with discrete multiple exposures}
			\begin{tabular}{|| c c ||}
				\hline
				Model    &  $f_{pop}(y)$  \\
				\hline 
				COM-Poisson     & $\sum_{k=0}^{T}\bigg[\frac{1}{Z(\theta_{t_k},\nu)}\frac{f_{t_k}(y)}{S_{t_k}(y)}\sum_{j=1}^{\infty}\frac{j\lbrace \theta_{t_k} S_{t_k}(y)\rbrace^j}{(j!)^v}\prod_{\stackrel{l\neq k}{l=0}}^{T}\frac{Z(\theta_{t_l}S_{t_l}(y),\nu)}{Z(\theta_{t_l},\nu)}\bigg]$    \\[2ex]
				Poisson & $\sum_{k=0}^{T}\lbrace\theta_{t_k}f_{t_k}(y)\rbrace\text{exp}\Big[\sum_{k=0}^{T}\lbrace-\theta_{t_k}F_{t_k}(y)\Big]$      \\ [2ex]
				Bernoulli    & $\sum_{k=0}^{T}\Big[\Big\{\frac{\theta_{t_k}}{1+\theta_{t_k}}f_{t_k}(y)\Big\}\prod_{\stackrel{l\neq k}{l=0}}^{T} \Big\{ \frac{1+\theta_{t_l}S_{t_l}(y)}{1+\theta_{t_l}} \Big\}\Big]$   \\ [2ex]
				Geometric & $\sum_{k=0}^{T}\Big[\frac{(1-\theta_{t_k})\theta_{t_k}f_{t_k}(y)}{(1-\theta_{t_k}S_{t_k}(y))^2}\prod_{\stackrel{l\neq k}{l=0}}^{T}\frac{1-\theta_{t_l}}{1-\theta_{t_l}S_{t_l}(y)}\Big]$\\ [2ex]
				\hline
			\end{tabular}
			\label{table:T2}
		\end{table}
	
	\section{Maximum likelihood estimation}
	
We develop the EM algorithm to carry out the maximum likelihood estimation of model parameters; see \citet{Pal16,Pal17b,Pal17a,Pal18c} and \citet{Jodi22}. First, let $F_{pop}$ and $F_1$ denote the cdf of the entire population and the susceptible population, respectively. Furthermore, let $S_1$ denote the survival function of the susceptible population. Define $J$ to be an indicator variable which takes on the value of 0 if the patient is immuned and 1 if the patient is susceptible. As such, we have $P[J=0]=p_0$ and $P[J=1]=1-p_0$. Furthermore, note that
	$$F_{pop}(y)=(1-p_0)F_1(y)$$
	and hence
	$$S_{pop}(y)=1-F_{pop}(y)=p_0+(1-p_0)S_1(y).$$
	Using the form of $S_{pop}(y)$ as in \eqref{Sp}, we can get an expression for $S_1$ as
	$$S_1(y)=\frac{S_{pop}(y)-p_0}{1-p_0},$$
	where $p_0$ is defined as in \eqref{p0}.
	 Let $\delta$ denote the censoring indicator such that when $\delta=1$, the actual time-to-event is observed and $\delta=0$ when the time-to-event is right censored. Let us define two sets, say, $I_1$ and $I_0$ as  $I_1=\lbrace i:\delta_i=1\rbrace$ and $I_0=\lbrace i:\delta_i=0\rbrace$. Note that for the set $I_0$, the value of $J$ is unknown (since a right censored patient may be susceptible to the event of interest), and this introduces the missing data. Let $\bm{\theta}$ denote the unknown parameter vector of our model. Then, the complete data likelihood function is given by
		$$L_c(\bm{\theta})=\prod_{I_1}\lbrace f_{pop}(y_i)\rbrace\prod_{I_0}\lbrace p_{0i}\rbrace^{1-J_i}\lbrace(1-p_{0i})S_1(y_i)\rbrace^{J_i}.$$
	Given the above likelihood function, we can write the complete data log-likelihood function as
	\begin{eqnarray}
			l_c(\bm{\theta})&=\sum_{I_1}\log\lbrace f_{pop}(y_i)\rbrace+\sum_{I_0}(1-J_i)\log\lbrace p_{0i}\rbrace+\sum_{I_0}J_i\log\lbrace(1-p_{0i})S_1(y_i)\rbrace.
			\label{lc}
	\end{eqnarray}	
Now that we have an expression for the complete data log-likelihood function, we can begin discussion on the development of the EM algorithm; see \citet{Mcl08}. In the expectation step (E-step), we calculate the expectation of the complete log-likelihood function with respect to the distribution of the unobserved $J_i'$s, given the model parameters and the observed data. We should note here that $J_i'$s are Bernoulli random variables and are linear in the complete data log-likelihood function. As such, at the $r^{th}$ iteration step, we simply need to calculate $\pi_i^{(r)}=E(J_i|\boldsymbol\theta^{(r)},\text{data})$, for $i\in I_0$, where $\boldsymbol\theta^{(r)}$ denotes the current parameter value at the $r^{th}$ iteration of the EM algorithm. Hence, for the $i^{th}$ censored observation, we can calculate $\pi_i^{(r)}$ as
		\begin{ceqn}
			\begin{align*}
			\pi_i^{(r)}&=P[J=1|Y_i>y_i;\boldsymbol\theta^{(r)}]\\
			&=\frac{P[Y_i>y_i|J_i=1]P[J_i=1]}{P[Y_i>y_i]}\Big|_{\boldsymbol\theta=\boldsymbol\theta^{(r)}}\\
			&=\frac{(1-p_{0i})S_1(y_i)}{S_{pop}(y_i)}\Big|_{\boldsymbol\theta=\boldsymbol\theta^{(r)}}\\
			&=\frac{S_{pop}(y_i)-p_{0i}}{S_{pop}(y_i)}\Big|_{\boldsymbol\theta=\boldsymbol\theta^{(r)}}\\
			&=1-\frac{p_{0i}}{S_{pop}(y_i)}\Big|_{\boldsymbol\theta=\boldsymbol\theta^{(r)}}.
			\end{align*}
		\end{ceqn}
Therefore, in the E-step, we only replace $J_i$ in \eqref{lc} with ${\pi}_i^{(r)}$ if the $i^{th}$ observation is censored. As done in \citet{Mcl08}, we will denote the conditional expectation of the complete data log-likelihood function at the $r^{th}$ iteration as $Q(\boldsymbol\theta,\bm{\pi}^{(r)})$, where $\bm{\pi}^{(r)}$ is the vector of ${\pi}_i^{(r)}$ values.

The next step in the EM algorithm is the maximization step (M-step). In this step, we maximize the $Q(\bm{\theta},\bm{\pi}^{(r)})$ function with respect to $\bm{\theta}$ over the parameter space $\bm{\Theta}$. In other words, we choose $\bm{\theta}^{(r+1)}$ to be a value of $\boldsymbol\theta\in\boldsymbol\Theta$ such that
	$$\bm{\theta}^{(r+1)} =  \operatorname*{arg\, max}_{\boldsymbol\theta}Q(\boldsymbol\theta,\bm{\pi}^{(r)});$$ see \citet{PalRoy21,PalRoy22,PalRoy23}.
	We continue to repeat this algorithmic process until some convergence criterion is satisfied, for example, $|\frac{\bm{\theta}^{r+1}-\bm{\theta}^r}{\bm{\theta}^r}|<\epsilon$, where $\epsilon$ is some pre-defined tolerance such as $\epsilon = 0.001$. 
	
\subsection{Expressions of the $Q$-function}

%For this purpose, we define the following expressions based on our parametric Weibull assumption:

	%\begin{ceqn}
			%\begin{align*}
			%F_{t_k}(y|\gamma_1,\gamma_2)&= \begin{cases}
			%1-e^{-((y-t_k)/\gamma_2)^{\gamma_1}} &, y-t_k>0 \\
			%0 &, y-t_k\leq 0. \\
			%\end{cases}\\
			%f_{t_k}(y|\gamma_1,\gamma_2)&= \begin{cases}
			%\frac{\gamma_1}{\gamma_2}\big(\frac{y-t_k}{\gamma_2}\big)^{\gamma_1-1}e^{-((y-t_k)/\gamma_2)^{\gamma_1}} &, y-t_k>0 \\
			%0 &, y-t_k\leq 0. \\
			%\end{cases}\\
			%S_{t_k}(y|\gamma_1,\gamma_2)&= \begin{cases}
			%\exp\lbrace-((y-t_k)/\gamma_2)^{\gamma_1}\rbrace &, y-t_k>0\\
			%1 &, y-t_k\leq 0. \\
			%\end{cases}	
		%\end{align*}
	%\end{ceqn}
We present the simplified expressions of the $Q$-functions for the COM-Poisson cure model and its special cases with discrete multiple exposures.

	\subsubsection{COM-Poisson cure rate model with discrete multiple exposures}
	The $Q$-function, $Q(\boldsymbol\theta,\boldsymbol\pi^{(k)})$, can be expressed as
	\begin{ceqn}
		\begin{align*}
			Q(\boldsymbol\theta,\boldsymbol\pi^{(k)})&=\sum_{i\in I_1}\log\bigg[\sum_{k=0}^{T}\bigg\{\frac{1}{Z(\theta_{it_k},\nu)}\frac{f_{t_k}(y_i)}{S_{t_k}(y_i)}\sum_{j=1}^{\infty}\frac{j\lbrace \theta_{it_k} S_{t_k}(y_i)\rbrace^j}{(j!)^v}\prod_{\stackrel{l\neq k}{l=0}}^{T}\frac{Z(\theta_{it_l}S_{t_l}(y_i),\nu)}{Z(\theta_{it_l},\nu)}\bigg\}\bigg]\\
			&-\sum_{i\in I_0}(1-\pi_i^{(k)})\log\bigg\{\prod_{k=0}^{T}Z(\theta_{it_k},\nu)\bigg\}\\
			&+\sum_{i\in I_0}\pi_i^{(k)}\log\bigg\{\prod_{k=0}^{T}\frac{Z(\theta_{it_k}S_{t_k}(y_i),\nu)}{Z(\theta_{it_k},\nu)}-\prod_{k=0}^{T}\frac{1}{Z(\theta_{it_k},\nu)}\bigg\},
		\end{align*}
	\end{ceqn}
	where
	$$\pi_i^{(k)}=1-\prod_{k=0}^T \frac{1}{Z(\theta_{it_k}S_{t_k}(y_i),\nu)}\Big|_{\bm{\theta}=\bm{\theta}^{(k)}}.$$
	
	\subsubsection{Poisson cure rate model with discrete multiple exposures}
	
	The $Q$-function, $Q(\boldsymbol\theta,\boldsymbol\pi^{(k)})$, can be expressed as
	\begin{ceqn}
		\begin{align*}
			Q(\boldsymbol\theta,\boldsymbol\pi^{(k)})&=\sum_{i\in I_1}\Big[\log\Big\{\sumk \theta_{it_k}f_{t_k}(y_i)\Big\}-\sumk\theta_{it_k}F_{t_k}(y_i)\Big]\\
			&-\sum_{i\in I_0}(1-\pi_i^{(k)})\Big\{\sumk\theta_{it_k}\Big\}+\sum_{i\in I_0}\pi_i^{(k)}\log\Big\{ e^{-\sum_{k=0}^T\theta_{it_k}F_{t_k}(y_i)} - e^{-\sum_{k=0}^T\theta_{it_k}}  \Big\},
		\end{align*}
	\end{ceqn}
	
	where
	$$\pi_i^{(k)}=1-\frac{e^{-\sum_{k=0}^T\theta_{it_k}}}{ e^{-\sum_{k=0}^T\theta_{it_k}F_{t_k}(y_i)} }\Big|_{\bm{\theta}=\bm{\theta}^{(k)}}.$$
	
	\subsubsection{Bernoulli cure rate model with discrete multiple exposures}
	 The $Q$-function, $Q(\boldsymbol\theta,\boldsymbol\pi^{(k)})$, can be expressed as
	 \begin{ceqn}
	 	\begin{align*}
	 		Q(\boldsymbol\theta,\boldsymbol\pi^{(k)})&=\sum_{i\in I_1}\log\Big[ \sum_{k=0}^{T}\Big[\Big\{\frac{\theta_{it_k}}{1+\theta_{it_k}}f_{t_k}(y_i)\Big\}\prod_{\stackrel{l\neq k}{l=0}}^{T} \Big\{ \frac{1+\theta_{it_l}S_{t_l}(y_i)}{1+\theta_{it_l}} \Big\}\Big] \Big]    \\
		 	&-\sum_{i\in I_0}(1-\pi_i^{(k)})\Big\{ \sum_{k=0}^T \log(1+\theta_{it_k})  \Big\} +\sum_{i\in I_0}\pi_i^{(k)}\log\Big[ \prod_{k=0}^{T} \Big\{ \frac{1+\theta_{it_k}S_{t_k}(y_i)}{1+\theta_{it_k}}   \Big\} - \prod_{k=0}^{T} \big\{\frac{1}{1+\theta_{it_k}}  \big\} \Big],
	 	\end{align*}
	 \end{ceqn}
	
	where 
	$$\pi_i^{(k)}=1 - \prod_{k=0}^T \Big\{\frac{1}{1+\theta_{it_k}S_{t_k}(y_i)}\Big\}\Big|_{\bm{\theta}=\bm{\theta}^{(k)}}.  $$

	\subsubsection{Geometric cure rate model with discrete multiple exposures}
	 The $Q$-function $Q(\boldsymbol\theta,\boldsymbol\pi^{(k)})$ can be expressed as
	 \begin{ceqn}
	 	\begin{align*}
		 	Q(\boldsymbol\theta,\boldsymbol\pi^{(k)})&=\sum_{i\in I_1}\log\Big[  \sum_{k=0}^{T}\Big[\frac{(1-\theta_{it_k})\theta_{it_k}f_{t_k}(y_i)}{(1-\theta_{it_k}S_{t_k}(y_i))^2}\prod_{\stackrel{l\neq k}{l=0}}^{T}\frac{1-\theta_{it_l}}{1-\theta_{it_l}S_{t_l}(y_i)}\Big]   \Big]\\
		 	&+\sum_{i\in I_0}(1-\pi_i^{(k)})\Big\{ \sum_{k=0}^T \log(1-\theta_{it_k})\Big\} \\
		 	& + \sum_{i\in I_0} \pi_i^{(k)} \log \Big[ \prod_{k=0}^{T}\frac{1-\theta_{it_k}}{1-\theta_{it_k}S_{t_k}(y_i)} -  \prod_{k=0}^{T}(1-\theta_{it_k}) \Big],
	 	\end{align*}
	 \end{ceqn}
	 
	 where
	 $$\pi_i^{(k)}=1-\prod_{k=0}^{T}(1-\theta_{it_k}S_{t_k}(y_i)) \Big|_{\bm{\theta}=\bm{\theta}^{(k)}}.$$

\section{Simulation study}

We shall now assume the pathogen promotion times to follow a Weibull distribution with probability density function given by
\begin{equation}
f(y) = \frac{\gamma_1}{\gamma_2} \bigg(\frac{y}{\gamma_2} \bigg)^{\gamma_1-1} e^{-\big(\frac{y}{\gamma_2}\big)^{\gamma_1}}, \ \ y>0, \gamma_1>0, \gamma_2 > 0. 
\label{Wei}
\end{equation}
Thus, we now have $\boldsymbol\gamma = (\gamma_1,\gamma_2)^\prime$. Note that one can also assume other parametric lifetime distributions here; see \citet{Bal12,Bal15,Bal15b}. One can also model these promotion times assuming a semi-parametric framework or a completely parametric framework with respect to the pathogen promotion times; see \citet{Bal16b} and \citet{Bal17}. 

\subsection{Model Fitting}

We now describe how to generate time-to-event data in the context of infectious diseases with multiple exposures. Using the generated data, we test the performance of the proposed EM algorithm through the calculated bias, root mean square (RMSE), and coverage probabilities of the asymptotic confidence intervals. For our simulation study, we consider the time to an outbreak of a nosocomial pulmonary infection since it easily incorporates a multiple exposure scenario; see \citet{Tou07}. It is known that tracheal intubation and mechanical ventilation are risk factors for nosocomial pulmonary infections. Furthermore, exposures to infection are multiple, for instance, the initial infection can be during intubation itself and successive exposures can be during daily secretion aspirations. We generate the intubation duration, $T$, from a Uniform(2, 30) distribution. We then generate random “time jumps" between each patient's exposure points to simulate time passing between each moment of exposure. To incorporate covariates into our simulation study, we consider a scenario where the immunological status ($x_{\text{imm}}$, which can be good or poor) of patients influence the infection rate related to tracheal tube placement and the protocol used for aspirations ($x_{\text{prot}}$, which can be A or B) influence the daily infection occasions. Thus, we have $\theta_{t_0} = \exp(\beta_0 + \beta_1 x_{\text{imm}})$ and $\theta_{t_k} = \exp(\beta_2 + \beta_3 x_{\text{prot}}), k=1,2,...,T$, where $x_{\text{imm}} = 0$ for patients with poor immunological status and $= 1$ otherwise. Similarly, $x_{\text{prot}} = 1$ for patients undergoing protocol A and $= 0$ otherwise. We consider different sample sizes, $n=200$ and $n=400$, to study the behavior of the model under small and large sample sizes. We also consider two different choices of Weibull parameters for eqns. \eqref{F} and \eqref{S} as $(\gamma_1,\gamma_2) = (2.5,2.5)$ and $(1.5,3.5)$. For the choice of regression parameters, we consider $(\beta_0,\beta_1,\beta_2,\beta_3)=(0.5,-1,-3,2)$. To generate the time-to-event, we carry out the following steps:
\begin{description}
\item (i) At each exposure time $t_k, k=0,1,...,T,$ we generate $M_{t_k}$ pathogens from the COM-Poisson distribution with parameters $\theta_{t_k}$ and $\nu$, for a given choice of the shape parameter $\nu$. We also consider the special cases of the COM-Poisson distribution.
\item (ii) At exposure time $t_k$, if $M_{t_k} > 1,$ we generate $\{Z_{1,t_k},Z_{2,t_k},...,Z_{M_{t_k},t_k}\}$ progression times from the Weibull distribution using eqn. \eqref{Wei}. However, if $M_{t_k} = 0,$ then, we let $Z_{0,t_k} = \infty$.
\item (iii) Let $W_{t_k} = \min\{ Z_{1,t_k},Z_{2,t_k},...,Z_{M_{t_k},t_k}\}$, which represents the time-to-event at exposure time $t_k$. If $M_{t_k}=0,$ we let $W_{t_k} = \infty$.
\item (iv) Define the final time-to-event, after taking into account all exposure times, as \\
$W = \min\{ W_{t_0},W_{t_1},...,W_{t_k},...,W_{t_T}\}$.
\item (v) Let $C$ denote the random censoring time generated from an exponential distribution with a suitable rate $\alpha$ to meet the desired censoring proportion. Furthermore, let $Y$ denote the observed time-to-event. Then, if $W=\infty,$ we set $Y=C$. If $W < \infty,$ we set $Y=\min\{ W,C\}$.
\item (vi) Let $\delta$ denote the right censoring indicator. Then, set $\delta = 1,$ if $Y=W,$ and set $\delta = 0,$ if $Y=C$.
\end{description}
For our simulation study, we choose the censoring rate $\alpha$ to be 0.10. To find an initial guess to start the iterative algorithm, we create an interval for each parameter by taking 15\% deviation off the true value and then select a value at random from the created interval. In Tables \ref{table:Ber} and \ref{table:Pois}, we present the model fitting results for the Bernoulli and Poisson cure rate models, respectively, with multiple exposures. We leave out the geometric model since the choice of the link function does not guarantee that $\theta_{t_k} < 1$,  which is required for the geometric model, as we have discussed before. It is clear that our proposed EM algorithm can retrieve the true parameters values quite accurately. The bias, standard error (SE) and RMSE of the estimates are small. Furthermore, they all decrease with an increase in sample size, which is what one would expect. The 95\% coverage probabilities are also close to the nominal level. In Table \ref{table:COM}, we present the results for the COM-Poisson model with multiple exposures. Note that in this case the shape parameter $\nu$ has to be estimated in addition to other model parameters. For estimating the parameter $\nu,$ we employ a profile likelihood technique within the EM algorithm, which is done along the lines of \citet{Bal16}. For this purpose, first, we select a set of admissible values of $\nu$. Then, for each chosen value of $\nu,$ we run the EM algorithm to estimate the other model parameters and calculate the log-likelihood value at the MLEs. Finally, we select the MLE of $\nu$ as the value of $\nu$ that results in the highest log-likelihood value. From Table \ref{table:COM}, it is clear that this profile likelihood approach performs well in estimating the parameter $\nu$. The estimates of other model parameters are also close to the their true values. Once again, the coverage probabilities are close to the nominal level used. 

%For the purpose of the maximum likelihood estimation, we will now re-parameterize our observed lifetimes to account for the time delay between each exposure time. We do this using the following formula:
%		$$z_{i,t_k}= \begin{cases}
%	Y_i-t_k & Y_i-t_k>0 \\
%	0 & Y_i-t_k\leq 0. \\
%	\end{cases}
%	$$
% We can then use these points for the estimation of the parameters in the model. As described previously, we need these time-adjusted points to take into account the amount of time that has passed between each exposure time.
		
\begin{table} [H]
	%\scriptsize
	\caption{Model fitting results for the Bernoulli cure rate model with multiple exposures.}
	\begin{tabular*}{\textwidth}{@{\extracolsep{\fill}} l l l l l l l}\\ 
		\hline                                 
		$n$ & Parameter & Estimate & SE & Bias & RMSE & CP 95\% \\
		\hline   
		
		400 & $\beta_0=0.5$  & 0.4770  & 0.3005  &-0.0230  & 0.2898  & 0.950\\
		& $\beta_1=-1$   &-1.0438  & 0.3191  &-0.0438  & 0.3225  & 0.940\\
		& $\beta_2=-3$   &-2.9912  & 0.2890  & 0.0088  & 0.2619  & 0.940\\
		& $\beta_3=	2$   & 2.0321  & 0.3336  & 0.0321  & 0.2695  & 0.970\\ 
		& $\gamma_1=2.5$ & 2.5554  & 0.2095  & 0.0554  & 0.2213  & 0.935\\
		& $\gamma_2=2.5$ & 2.4398  & 0.1608  & 0.0602  & 0.1457  & 0.965\\ [2ex]
		%\hline
		
		200  & $\beta_0=0.5$  & 0.5270  & 0.4168  & 0.0270  & 0.3982  & 0.940\\
		& $\beta_1=-1$   &-1.0379  & 0.4649  &-0.0379  & 0.4955  & 0.935\\
		& $\beta_2=-3$   &-3.1321  & 0.4935  &-0.1321  & 0.7061  & 0.960\\
		& $\beta_3=2$    & 2.2151  & 0.5552  & 0.2151  & 0.7774  & 0.930\\
		& $\gamma_1=2.5$ & 2.5889  & 0.2852  & 0.0889  & 0.3335  & 0.925\\
		& $\gamma_2=2.5$ & 2.4983  & 0.2096  & 0.0017  & 0.1948  & 0.940\\ [2ex]
		%\hline
		400  & $\beta_0=0.5$  & 0.4155  & 0.4383  &-0.0845  & 0.4395  & 0.950\\
		& $\beta_1=-1$   &-1.0541  & 0.3935  &-0.0541  & 0.3331  & 0.975\\
		& $\beta_2=-3$   &-2.9738  & 0.3831  & 0.0262  & 0.3973  & 0.930\\
		& $\beta_3=2$    & 2.9755  & 0.4174  &-0.0245  & 0.4588  & 0.915\\ 
		& $\gamma_1=1.5$ & 1.5570  & 0.1274  & 0.0570  & 0.1359  & 0.965\\
		& $\gamma_2=3.5$ & 3.3984  & 0.3799  &-0.1016  & 0.4136  & 0.905\\ [2ex]
		%\hline
		
		200  & $\beta_0=0.5$   & 0.4308  & 0.6067  &-0.0692  & 0.5680  & 0.935\\
		& $\beta_1=-1$    &-0.9745  & 0.5598  & 0.0255  & 0.5332  & 0.950\\
		& $\beta_2=-3$    &-2.9882  & 0.5582  & 0.0118  & 0.4703  & 0.920\\
		& $\beta_3=2$     & 1.9906  & 0.6054  &-0.0094  & 0.5151  & 0.910\\
		& $\gamma_1=1.5$  & 1.5788  & 0.1911  & 0.0788  & 0.2375  & 0.935\\
		& $\gamma_2=3.5$  & 3.3074  & 0.5371  &-0.1916  & 0.5426  & 0.895\\[2ex]
		%\hline
		%\hline
		\hline
	\end{tabular*}
	\label{table:Ber}
\end{table}
		
\begin{table} [H]
	%\scriptsize
	\caption{Model fitting results for the Poisson cure rate model with multiple exposures.}
	\begin{tabular*}{\textwidth}{@{\extracolsep{\fill}} l l l l l l l}\\ 
		\hline                                 
		$n$ & Parameter & Estimate & SE & Bias & RMSE & CP 95\% \\
		\hline

		400  & $\beta_0=0.5$  & 0.5060  & 0.1262  & 0.0060  & 0.1239  & 0.955\\
		& $\beta_1=-1$   &-1.0137  & 0.1605  &-0.0137  & 0.1757  & 0.935\\
		& $\beta_2=-3$   &-3.0621  & 0.4553  &-0.0621  & 0.4644  & 0.960\\
		& $\beta_3=	2$   & 2.0614  & 0.4936  & 0.0614  & 0.5037  & 0.965\\ 
		& $\gamma_1=2.5$ & 2.5282  & 0.1654  & 0.0282  & 0.1687  & 0.960\\
		& $\gamma_2=2.5$ & 2.4936  & 0.1604  &-0.0064  & 0.1577  & 0.965\\ [2ex]
		%\hline
		
		200  & $\beta_0=0.5$  & 0.5110  & 0.1785  & 0.0110  & 0.1713  & 0.970\\
		& $\beta_1=-1$   &-1.0370  & 0.2303  &-0.0370  & 0.2326  & 0.965\\
		& $\beta_2=-3$   &-3.0768  & 0.4916  &-0.0768  & 0.4923  & 0.965\\
		& $\beta_3=2$    & 2.0866  & 0.5459  & 0.0866  & 0.5317  & 0.955\\
		& $\gamma_1=2.5$ & 2.5326  & 0.2329  & 0.0326  & 0.3536  & 0.945\\
		& $\gamma_2=2.5$ & 2.5180  & 0.2264  & 0.0180  & 0.2318  & 0.930\\ [2ex]
		%\hline
		
		400  & $\beta_0=0.5$  & 0.5069  & 0.1867  & 0.0069  & 0.1818  & 0.960\\
		& $\beta_1=-1$   &-1.1014  & 0.1817  &-0.1014  & 0.1823  & 0.945\\
		& $\beta_2=-3$   &-3.1269  & 0.6654  &-0.1269  & 0.6932  & 0.900\\
		& $\beta_3=2$    & 2.1127  & 0.6983  & 0.1127  & 0.7281  & 0.910\\ 
		& $\gamma_1=1.5$ & 1.5155  & 0.1062  & 0.0155  & 0.1033  & 0.965\\
		& $\gamma_2=3.5$ & 3.6233  & 0.4127  & 0.1233  & 0.4266  & 0.905\\[2ex]
		%\hline
		
		200  & $\beta_0=0.5$   & 0.4729  & 0.2687  &-0.0281  & 0.2744  & 0.945\\
		& $\beta_1=-1$    &-1.0477  & 0.2688  & 0.0477  & 0.2653  & 0.965\\
		& $\beta_2=-3$    &-3.0373  & 0.7598  &-0.0373  & 0.5923  & 0.920\\
		& $\beta_3=2$     & 2.0541  & 0.8183  & 0.0541  & 0.6590  & 0.915\\
		& $\gamma_1=1.5$  & 1.5466  & 0.1579  & 0.0466  & 0.1769  & 0.930\\
		& $\gamma_2=3.5$  & 3.4580  & 0.8309  & 0.0420  & 0.8945  & 0.905\\[2ex]
		%\hline
		
		\hline
	\end{tabular*}
	\label{table:Pois}
\end{table}

\begin{table} [H]
	%\scriptsize
	\caption{Model fitting results for the COM-Poisson cure rate model with multiple exposures (to apply the profile likelihood, the set of values of $\nu$ is chosen as $\{1.6,1.7,...,2.4\}$).}
	\begin{tabular*}{\textwidth}{@{\extracolsep{\fill}} l l l l l l l}\\ 
		\hline                                 
		$n$ & Parameter & Estimate & SE & Bias & RMSE & CP 95\% \\
		\hline   
		
		400  & $\beta_0=0.5$  & 0.4644  & 0.2259  &-0.0356  & 0.2282  & 0.945\\
		& $\beta_1=-1$   &-1.0151  & 0.2345  &-0.0151  & 0.2344  & 0.940\\
		& $\beta_2=-3$   &-3.0504  & 0.2121  &-0.0504  & 0.2175  & 0.960\\
		& $\beta_3=	2$   & 2.0522  & 0.2177  & 0.0522  & 0.2233  & 0.960\\ 
		& $\gamma_1=2.5$ & 2.5842  & 0.2575  & 0.0842  & 0.2703  & 0.965\\
		& $\gamma_2=2.5$ & 2.5376  & 0.2569  & 0.0376  & 0.2590  & 0.955\\
		&$\nu=2$		 & 1.8320  & --		 &-0.1680  & 0.8133  & --\\[2ex]
		%\hline
		
		200  & $\beta_0=0.5$  & 0.4554  & 0.5288  &-0.0446  & 0.5280  & 0.940\\
		& $\beta_1=-1$   &-1.1001  & 0.4805  &-0.1001  & 0.5149  & 0.950\\
		& $\beta_2=-3$   &-3.0033  & 0.3293  &-0.0033  & 0.3277  & 0.960\\
		& $\beta_3=2$    & 1.9846  & 0.3507  &-0.0126  & 0.3492  & 0.960\\
		& $\gamma_1=2.5$ & 2.6991  & 0.4356  & 0.1991  & 0.4770  & 0.925\\
		& $\gamma_2=2.5$ & 2.5084  & 0.3260  & 0.0084  & 0.3345  & 0.920\\ 
		& $\nu=2$ 		 & 1.7975  & -- 	  &-0.2025  & 0.8956  & --\\[2ex]
		%\hline
		
		400  & $\beta_0=0.5$  & 0.4470  & 0.3462  &-0.0630  & 0.3485  & 0.950\\
		& $\beta_1=-1$   &-1.0812  & 0.4155  &-0.0812  & 0.4213  & 0.945\\
		& $\beta_2=-3$   &-3.0359  & 0.2481  &-0.0359  & 0.2494  & 0.910\\
		& $\beta_3=2$    & 2.0275  & 0.2792  & 0.0275  & 0.2791  & 0.920\\ 
		& $\gamma_1=1.5$ & 1.5556  & 0.1629  & 0.0556  & 0.1714  & 0.945\\
		& $\gamma_2=3.5$ & 3.6636  & 0.5117  & 0.1636  & 0.5348  & 0.950\\ 
		& $\nu=2$ 		 & 2.1878  & --      & 0.1878  & 0.8401  & --\\[2ex]
		%\hline
		
		200  & $\beta_0=0.5$   & 0.5433  & 0.5760  & 0.0433  & 0.5746  & 0.925\\
		& $\beta_1=-1$    &-1.1837  & 0.6957  &-0.1837  & 0.7579  & 0.945\\
		& $\beta_2=-3$    &-2.9712  & 0.3775  & 0.0288  & 0.3766  & 0.900\\
		& $\beta_3=2$     & 1.9377  & 0.3555  &-0.0623  & 0.3591  & 0.915\\
		& $\gamma_1=1.5$  & 1.5665  & 0.2647  & 0.0665  & 0.2716  & 0.935\\
		& $\gamma_2=3.5$  & 3.5186  & 0.7116  & 0.0186  & 0.7082  & 0.940\\
		& $\nu=2$ 		  & 1.7864  & --      &-0.2136  & 0.9376  & --\\[2ex]
		%\hline
		\hline
	\end{tabular*}
	\label{table:COM}
\end{table}

\subsection{Model Discrimination}
	
Due to the flexibility of the COM-Poisson cure rate model with discrete multiple exposures and its inclusion of several other multiple exposure cure rate models as special cases, we are in a position to select a simple multiple exposure cure rate model within the bigger family of COM-Poisson multiple exposure cure rate model that provides an adequate fit as the COM-Poisson multiple exposure model itself in many cases. This motivates us to explore the flexibility of the proposed multiple exposure COM-Poisson cure rate model to select a parsimonious cure rate model that provides an adequate fit to the given data. To this end, we carry out two different model discrimination studies, one using the likelihood ratio test and the other using the information-based criteria.

\subsubsection{Likelihood ratio test}

In this model discrimination study, we investigate the performance of the likelihood ratio test in testing the null hypothesis that the distribution of the number of pathogens at each exposure time can be described by one of the Bernoulli $(\nu\rightarrow\infty)$, Poisson $(H_0:\nu=1)$, COM-Poisson $(\nu=0.5)$, and COM-Poisson $(\nu=2)$ distributions versus the alternative hypothesis that the number of pathogens can be described by any other member of the COM-Poisson family besides the one already specified in the null hypothesis. The likelihood test statistic is defined as $\Lambda=-2(\hat{l}_0-\hat{l})$, where $\hat{l}_0$ denotes the maximized log-likelihood function value under the null hypothesis and $\hat{l}$ denotes the unrestricted maximized log-likelihood function value. Note that to calculate $\hat{l}$, we fit the multiple exposure COM-Poisson model for which the profile likelihood technique needs to be employed. For this simulation study, we consider the following three parameter settings: (i) Setting 1 considers 400 patients with parameters $(\beta_0,\beta_1,\beta_2,\beta_3,\gamma_1,\gamma_2)=(0.5,-1,-3,2,2.5,2.5)$; (ii) Setting 2 considers 400 patients with parameters $(\beta_0,\beta_1,\beta_2,\beta_3,\gamma_1,\gamma_2)=(0.5,-1,-3,2,1.5,3.5)$; and (iii) Setting 3 considers 200 patients with parameters $(\beta_0,\beta_1,\beta_2,\beta_3,\gamma_1,\gamma_2)=(0.5,-1,-3,2,2.5,2.5)$. For each simulated data from a true model, we calculate the likelihood ratio test statistic of the fitted Bernoulli, Poisson, COM-Poisson ($\nu=0.5$), and COM-Poisson ($\nu=2$) models versus the fitted COM-Poisson model. Based on 200 data sets for each true model and for each parameter setting, and using 10\% level of significance, we report the observed significance levels (in bold) and observed power values of the likelihood ratio test in Table \ref{table:LR}. These values are obtained by the rejection rates of the null hypotheses. From Table \ref{table:LR}, it can be seen that the asymptotic null distribution of the likelihood ratio test statistic is reasonably approximated. When the true model is Bernoulli (or COM-Poisson ($\nu=0.5$)), the power to reject the COM-Poisson ($\nu=0.5$) (or Bernoulli) is high. Thus, the likelihood ratio test can discriminate between the Bernoulli and COM-Poisson ($\nu=0.5$) models. In this regard, note that the rejection rate is higher when the true model is COM-Poisson ($\nu=0.5$) and the fitted model is Bernoulli. Now, when the true model is Poisson, the likelihood ratio test still possess adequate power to reject the Bernoulli model. However, when the true model is Bernoulli, the test has very low power to reject the Poisson model. Finally, the power of the likelihood ratio test to discriminate among COM-Poisson ($\nu=0.5$), Poisson, and COM-Poisson ($\nu=2$) models vary from low to medium. 

\begin{table} [ht!]
	%\small
	\caption{Observed levels and observed power values of the likelihood ratio test.}
	\begin{tabular*}{\textwidth}{@{\extracolsep{\fill}} l l l l l} 
		\hline                                 
		&\multicolumn{4}{c}{True multiple exposure model } \\  \cline{2-5}
		Fitted model & $\nu=0.5$ & $\nu=1$ & $\nu=2$ & $\nu\rightarrow\infty$\\
		\hline
		& \multicolumn{4}{c}{Setting 1}\\
		$\nu=0.5$   			& $\boldsymbol{0.100}$ & 0.315 & 0.465 & 0.765 \\
		$\nu=1$     			& 0.230 & $\boldsymbol{0.095}$ & 0.105 & 0.140 \\
		$\nu=2$     			& 0.425 & 0.390 & $\boldsymbol{0.100}$ & 0.095 \\
		$\nu\rightarrow\infty$  & 0.945 & 0.745 & 0.400 & $\boldsymbol{0.070}$ \\ [1ex]
		
		& \multicolumn{4}{c}{Setting 2}\\
		$\nu=0.5$  				& $\boldsymbol{0.115}$ & 0.250 & 0.400 & 0.695 \\
		$\nu=1$   				& 0.205 & $\boldsymbol{0.125}$ & 0.085 & 0.155 \\
		$\nu=2$     			& 0.435 & 0.350 & $\boldsymbol{0.160}$ & 0.105 \\
		$\nu\rightarrow\infty$  & 0.885 & 0.715 & 0.410 &  $\boldsymbol{0.085}$ \\ [1ex]
		
		& \multicolumn{4}{c}{Setting 3}\\
		$\nu=0.5$   			& $\boldsymbol{0.145}$ & 0.225 & 0.380 & 0.680 \\
		$\nu=1$  				& 0.180 & $\boldsymbol{0.150}$ & 0.075 & 0.125 \\
		$\nu=2$     			& 0.375 & 0.355 & $\boldsymbol{0.170}$ & 0.090 \\
		$\nu\rightarrow\infty$  & 0.845 & 0.650 & 0.395 &  $\boldsymbol{0.065}$ \\ [1ex]
		\hline
	\end{tabular*}
	\label{table:LR}
\end{table}

\subsubsection{Information-based criteria}

In this model discrimination study, we investigate the performance of the Akaike information criterion (AIC) and the Bayesian information criterion (BIC) in choosing either the Bernoulli multiple exposure model, Poisson multiple exposure model, COM-Poisson $(\nu=0.5)$ multiple exposure model or COM-Poisson $(\nu=2)$ multiple exposure model, for a given true multiple exposure model. Our selected models cover both over-dispersed and under-dispersed models. We choose to look into the AIC and BIC since they are they two most widely used model selection criteria. The AIC is defined as $AIC=-2l+2p,$ where $l$ is the maximized log-likelihood value of the given model and $p$ is the number of parameters of the fitted model. The BIC, on the other hand, is defined as $BIC=-2l+p\log(n)$. Similar to the AIC, $l$ is the maximized log-likelihood value of the given model, $p$ is the number of parameters of the fitted model, and $n$ is the sample size. In using the AIC and BIC, the preferred model is the one with the lowest AIC or BIC value. 

For each generated true model, we fit all candidate models, i.e., Bernoulli $(\nu\rightarrow\infty)$, Poisson $(\nu=1)$, COM-Poisson ($\nu=0.5$), and COM-Poisson ($\nu=2$), and allow AIC/BIC to select the best model. To generate the true model, we consider the same parameter settings as in the case of likelihood ratio test. Based on 200 generated data sets for each true model and for each parameter setting, we calculate the observed selection rates for both AIC and BIC, and report these values in Table \ref{table:AIC}. Note that the selection rates for BIC turned out to be the same as that for AIC, and, as such, are not reported. From the results in Table \ref{table:AIC}, it is clear that the model selection criteria performs well in selecting the correct model. When the true model is Bernoulli (or COM-Poisson $(\nu = 0.5)$), the selection rate for COM-Poisson $(\nu = 0.5)$ (or Bernoulli) is very low. This suggests that the AIC and BIC can distinctly discriminate between these two models. Thus, the decision reached by AIC/BIC is in agreement with that reached by likelihood ratio test earlier. A similar conclusion can also be drawn when discriminating between Bernoulli and COM-Poisson $(\nu = 2)$ as well as between Bernoulli and Poisson models. However, the discrimination power of AIC and BIC among COM-Poisson $(\nu = 0.5)$, Poisson and COM-Poisson $(\nu = 2)$ models appear to be weak. Note, in this regard, that when the true model is COM-Poisson $(\nu = 2),$ the selection rate for COM-Poisson $(\nu = 0.5)$ is low. These observations suggest that it is worth exploring the flexibility of the COM-Poisson distribution to select a suitable distribution for the count on the pathogens at each exposure time. We note the advantage of using AIC/BIC in terms of speed. Unlike the method of model discrimination using likelihood ratio test, we do not need to estimate the shape parameter $v$ since we are only attempting to fit the candidate models for which the values of $\nu$ are specified. If we were to estimate the shape parameter $\nu$, we would need to use a profile likelihood approach, which would drastically increase the computation time. 

\begin{table} [ht!]
	\caption{Observed selection rates based on AIC}
	\begin{tabular*}{\textwidth}{@{\extracolsep{\fill}}  l l l l l}\\ 
		\hline                               
		&\multicolumn{4}{c}{True multiple exposure model } \\  \cline{2-5}
		Fitted model &  $\nu=0.5$ & $\nu=1$ & $\nu=2$ & $\nu\rightarrow\infty$\\
		\hline
		& \multicolumn{4}{c}{Setting 1}\\[0.5ex]
		$\nu=0.5$    			& 0.405  & 0.275  & 0.150  & 0.035 \\
		$\nu=1$    				& 0.310  & 0.410  & 0.315  & 0.105 \\ 
		$\nu=2$      			& 0.235  & 0.215  & 0.330  & 0.145 \\
		$\nu\rightarrow\infty$  & 0.050  & 0.100  & 0.205  & 0.715 \\[1ex]
		
		& \multicolumn{4}{c}{Setting 2}\\[0.5ex]
		$\nu=0.5$    			& 0.415  & 0.290  & 0.165  & 0.060 \\
		$\nu=1$    				& 0.320  & 0.425  & 0.320  & 0.110 \\ 
		$\nu=2$      			& 0.215  & 0.200  & 0.325  & 0.125 \\
		$\nu\rightarrow\infty$  & 0.050  & 0.085  & 0.190  & 0.705 \\[1ex]
		& \multicolumn{4}{c}{Setting 3}\\[0.5ex]
		$\nu=0.5$     			& 0.420  & 0.285  & 0.140  & 0.025 \\
		$\nu=1$    				& 0.300  & 0.400  & 0.295  & 0.140 \\ 
		$\nu=2$      			& 0.235  & 0.220  & 0.340  & 0.165 \\
		$\nu\rightarrow\infty$  & 0.045  & 0.095  & 0.225  & 0.670 \\[1ex]
		\hline
	\end{tabular*}
	\label{table:AIC}
\end{table}

\section{Real data application}

\begin{figure}[htb!]
\centering
\includegraphics[width=0.8\linewidth]{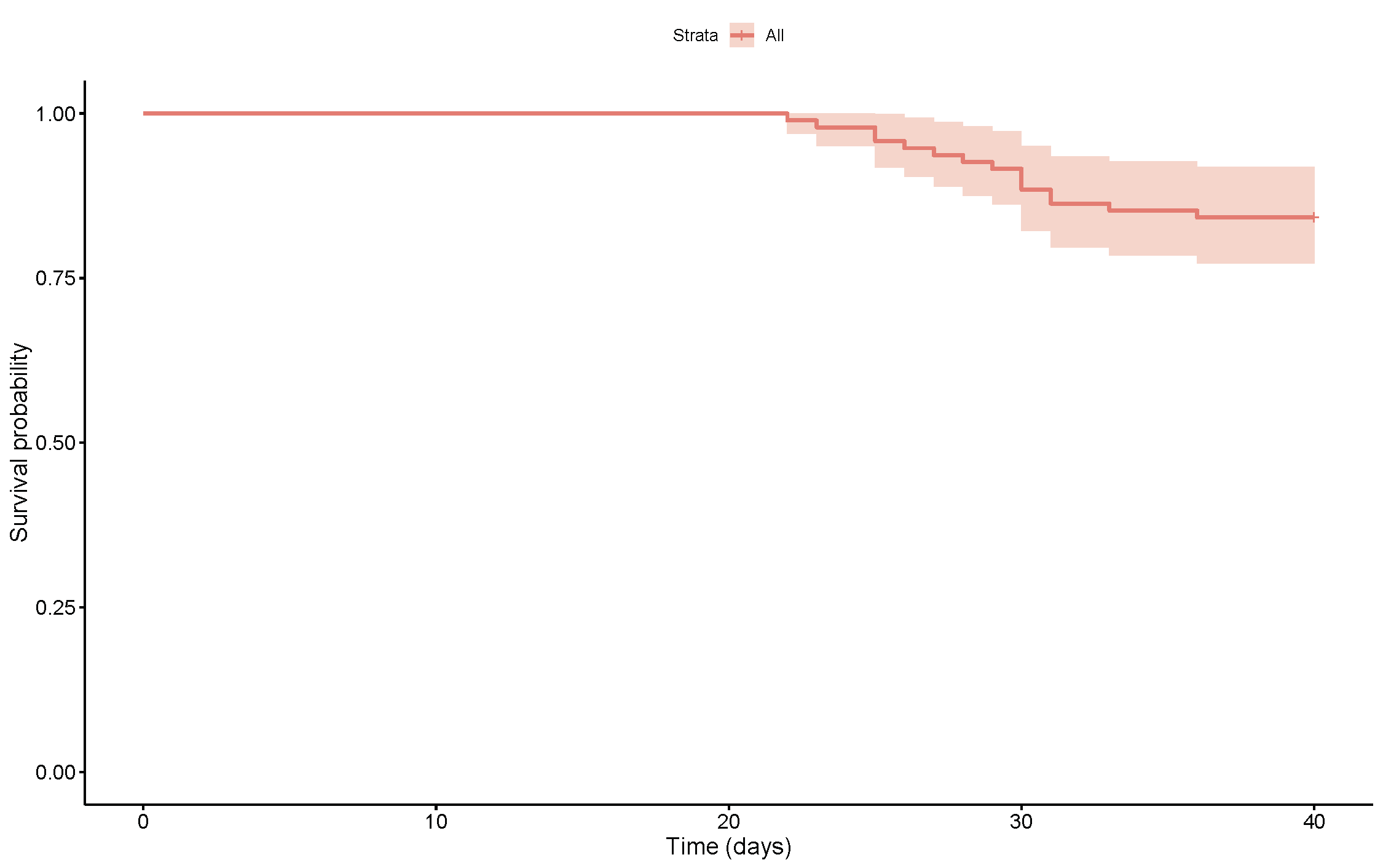}
\caption{Kaplan Meier estimates of the survival probability}
\label{figure:F1}
\end{figure}

We analyzed a data that can be found via the website \textit{Kaggle}, which is a free-to-access online database repository that contains thousands of data sets created by researchers worldwide. The data was collected by the World Health Organization (WHO) in conjunction with John Hopkins University on December 31, 2019. This data tracks patients who have contracted the SARS-CoV-2, also known as COVID-19 or simply coronavirus, after visiting Wuhan City, Hubei Province of China, which is believed to the epicenter of the COVID-19 pandemic. This data set contains the patient data of people who have been in contact with people from Wuhan or someone who had recently visited Wuhan. The data set has information on variables such as country of origin, gender, age, date of symptom onset, if the patient died, and if the patient recovered. Most importantly, the data set contains the day when the patient started to be exposed to COVID-19 and when the exposure ended, which is crucial for our model. However, not all patients had this information. As such, we had to comb through the data to find patients who had the values we desired, so we could create a new data set with the information we required. We began by selecting patients who had a definitive exposure start date and exposure end date. This left us with 120 patients to study. We decided the event of interest we wanted to study would be the time to recovery. From these patients, we decided to study the effects of the covariates age and gender on the recovery time of the patients. Once we eliminated patients whose age and gender were not available, we were left with 95 patients whose starting exposure time, ending exposure time, age, and gender were recorded. Of these 95 patients, 15 of them had recovered from the disease and their date of recovery was also recorded. Figure \ref{figure:F1} presents the Kaplan Meier (KM) estimates of the survival probability. It is clear that the KM curve levels off to a significant non-zero proportion, which indicates the presence of a cured sub-group. Hence, the proposed model is appropriate for this dataset. 

\subsection{Data preparation for model}

To begin, we identify our two covariates of interest as age and gender denoted $X_{age}$ and $X_{gen}$, respectively. We will treat gender as a binary covariate where $X_{gen}=1$ if the patient is male (54.7\%) and $X_{gen}=0$ if the patient is female (45.3\%). Next, we will let our covariate age be a categorical covariate. For this purpose, we divide the patients into 5 groups that were representative of the dispersion of the age. $X_{age}=1$ if the patient is between 0 and 24.5 years old (8.4\%), $X_{age}=2$ if the patient is between 24.5 and 36.5 years old (23.2\%), $X_{age}=3$ if the patient is between 36.5 and 48.5 years old (29.5\%), $X_{age}=4$ if the patient is between 48.5 and 60.5 years old (25.3\%), and $X_{age}=5$ if the patient is older than 60.5 years old (13.7\%). Now that we have defined our covariates and their values, we can use the log-linear link function as $\theta_{t_k}=\exp(\beta_0+\beta_1X_{gen}+\beta_2X_{age})$. Since we have no evidence of heterogeneity of exposure intensity with respect  to moment of exposure, we will assume $\theta_{t_k}$ is the same for each moment of exposure for each patient. Since the data records the number of days each patient was exposed to COVID-19, we will let the number of exposures be the number of days the patient was exposed. Therefore, the ``time jump" between each moment of exposure will be 1. The average number of exposures is 6.24, with a minimum value of 1 and a maximum value of 29. Furthermore, we will let the value for the days until recovery, in other words the time to the event of interest, to be the number of days that have passed from the first day of exposure until the recovery date. The average time to recovery is 29.125 days, the minimum value is 22 days, and the maximum value is 40 days. If the recovery time of the patient is not recorded, the patient's time to event is censored. The dataset has 85\% censored observations. Next, to begin the iterative process of the EM algorithm, an initial guess of parameters is needed. To find the initial guess of the parameters in the Weibull distribution, we found the mean and variance of the recorded lifetimes. We then used the known expressions for the mean and variance of the Weibull distribution and solved for the progression time parameters. To get an initial guess for the regression coefficients, we performed a grid search using the observed likelihood function. For the purpose of the grid search, we assumed the data to follow the Poisson model. We then set the values for the lifetime parameters as constant. Finally, we calculated the log-likelihood values using the constant lifetime parameters, and different combinations of the regression coefficients with values ranging from $\lbrack-5,5\rbrack$. Once we found the maximum log-likelihood value, we used those chosen parameters that resulted in the maximum log-likelihood value as the initial guess to begin the iterative process of the COM-Poisson model. We used these initial guesses and different values of $\nu$ to determine the correct model for this data. We selected values of $\nu$ ranging from 0 to 2, with jumps of 0.1, as well as the Bernoulli case ($\nu\rightarrow\infty$). 

\subsection{Real data results}

\begin{figure}[htb!]
\centering
\includegraphics[width=0.8\linewidth]{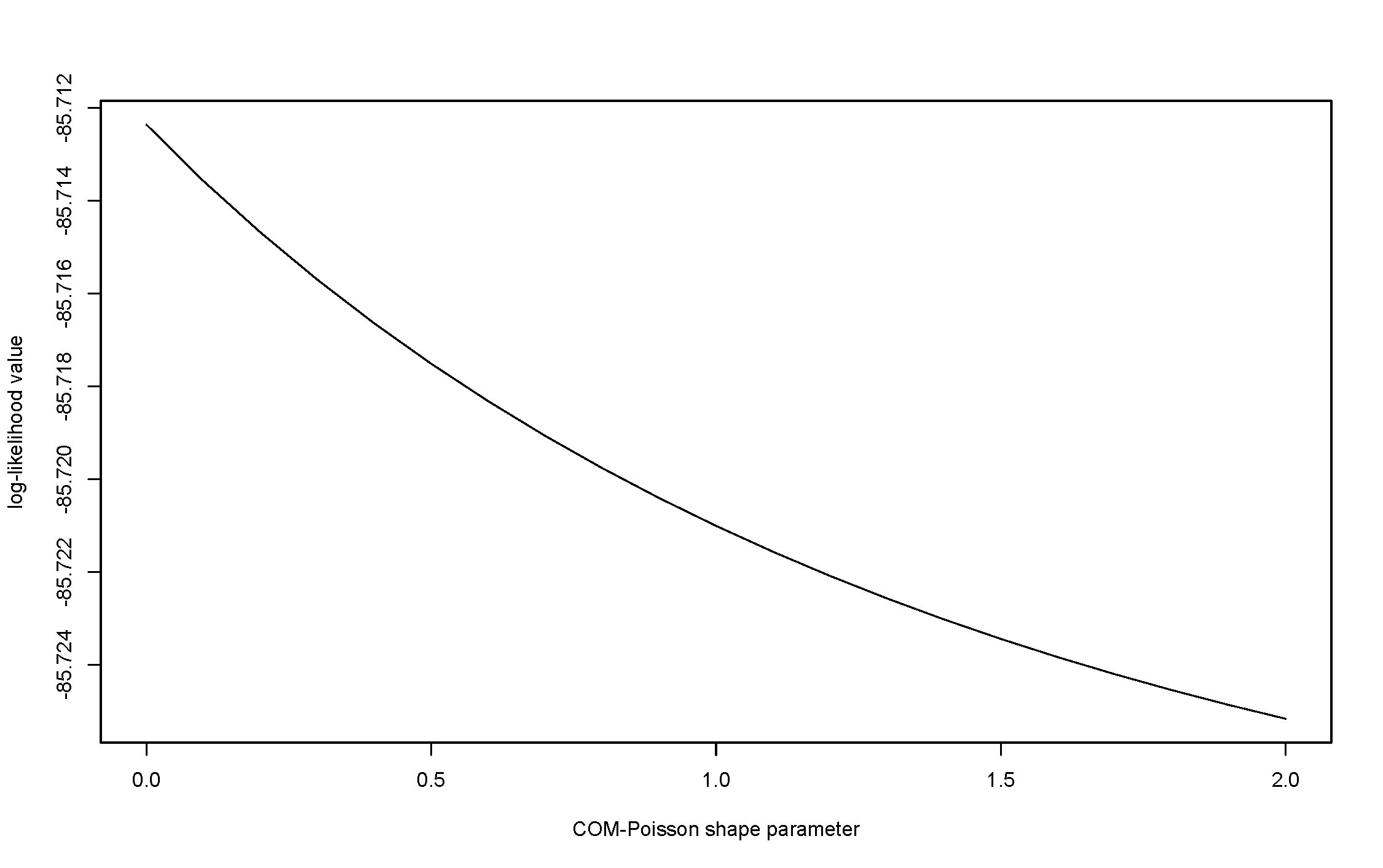}
\caption{Plot of the maximized log-likelihood value against the COM-Poisson shape parameter ($\nu$)}
\label{figure:F2}
\end{figure}

\begin{figure}[hptb!]
\centering
\begin{tabular}{ccc}
		\includegraphics[scale=0.30]{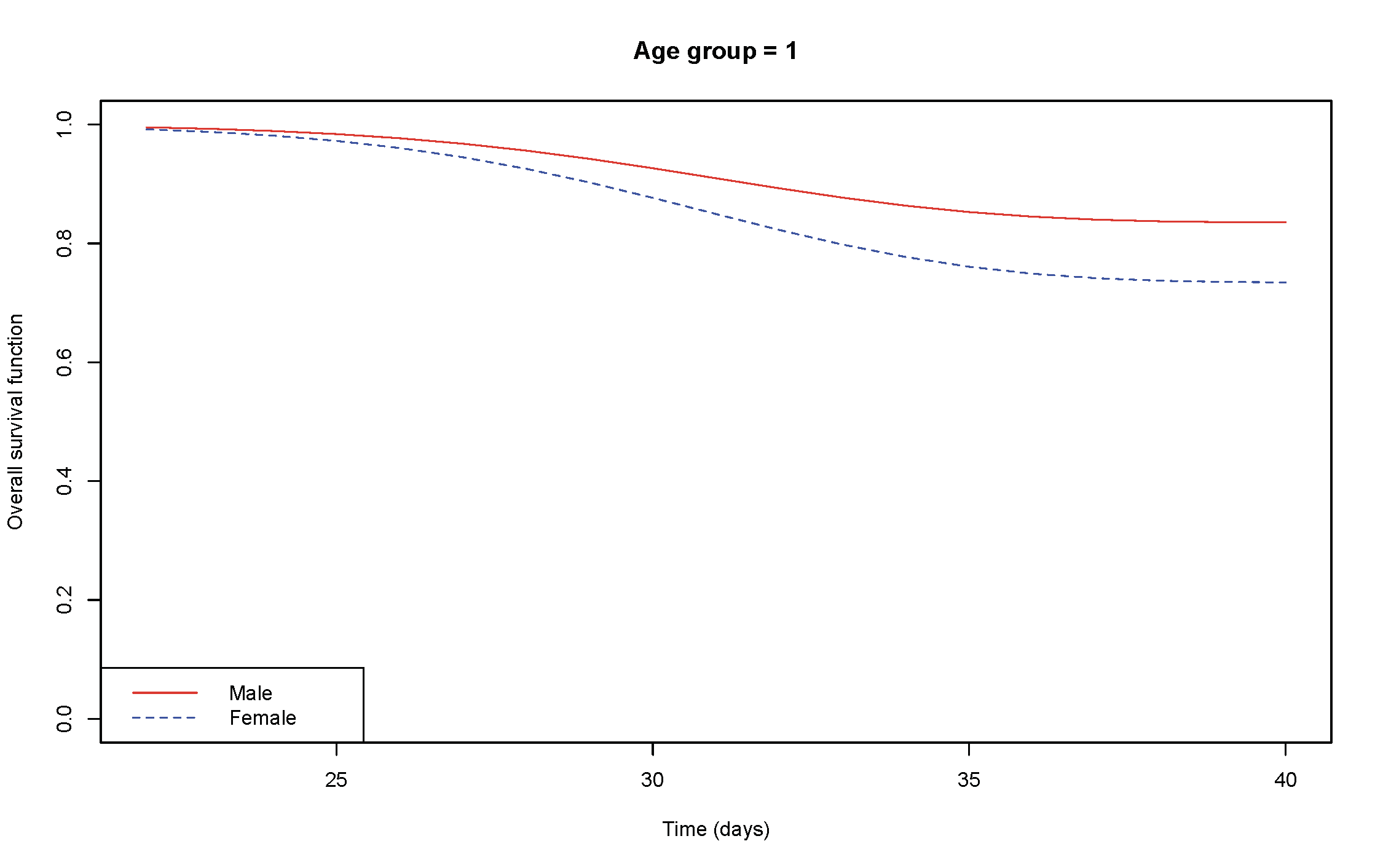}
		\includegraphics[scale=0.30]{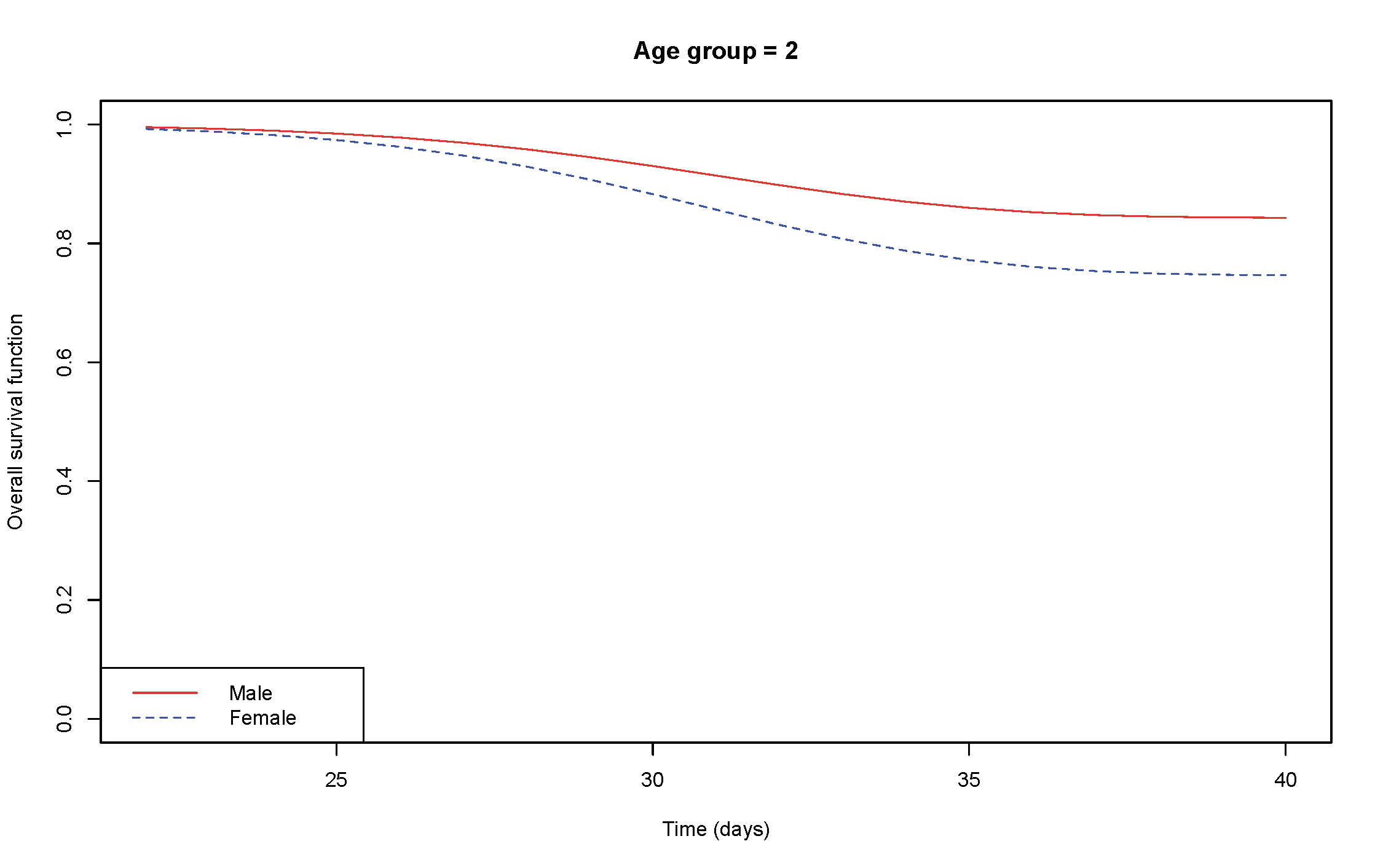}\\
		\includegraphics[scale=0.30]{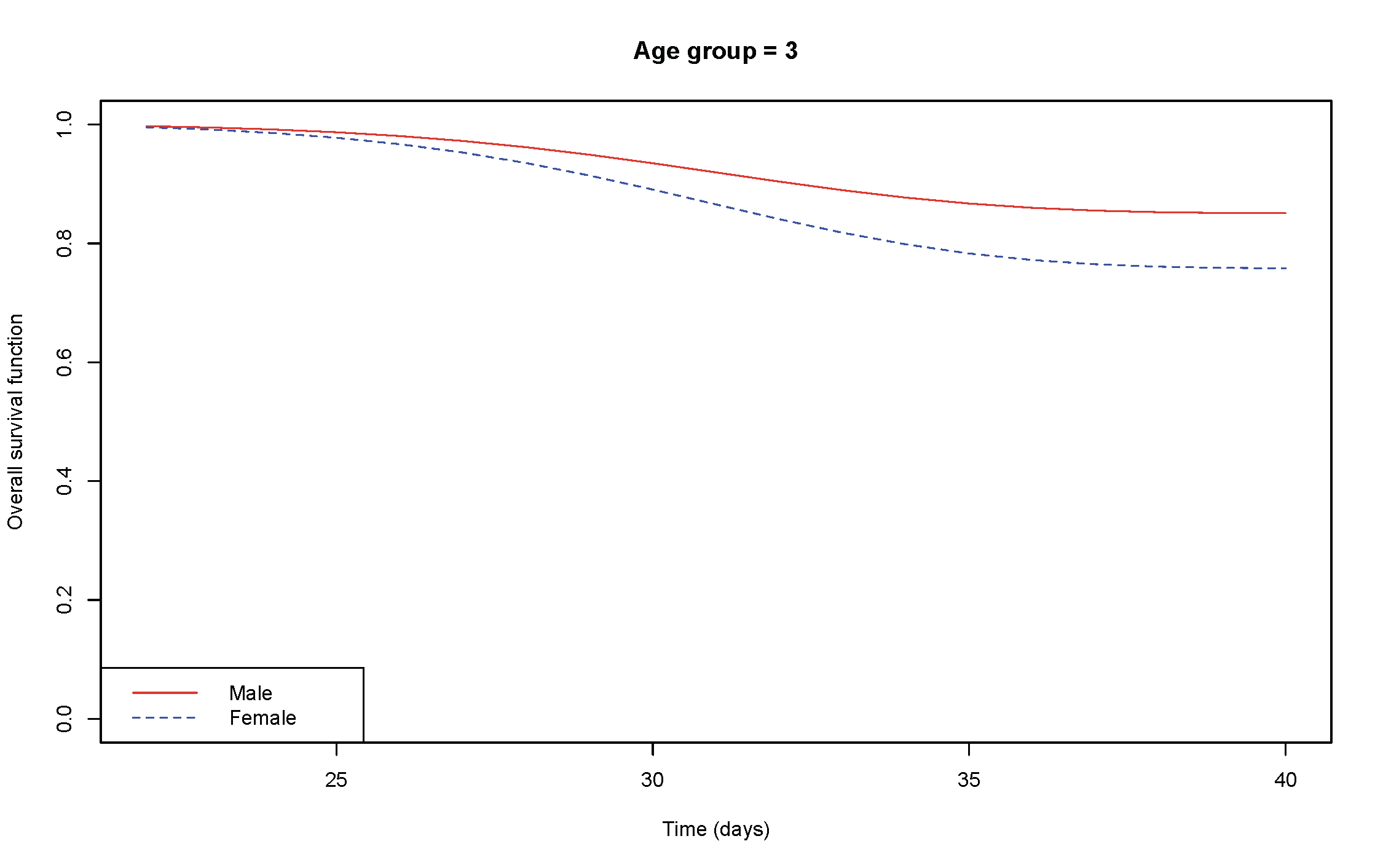}
		\includegraphics[scale=0.30]{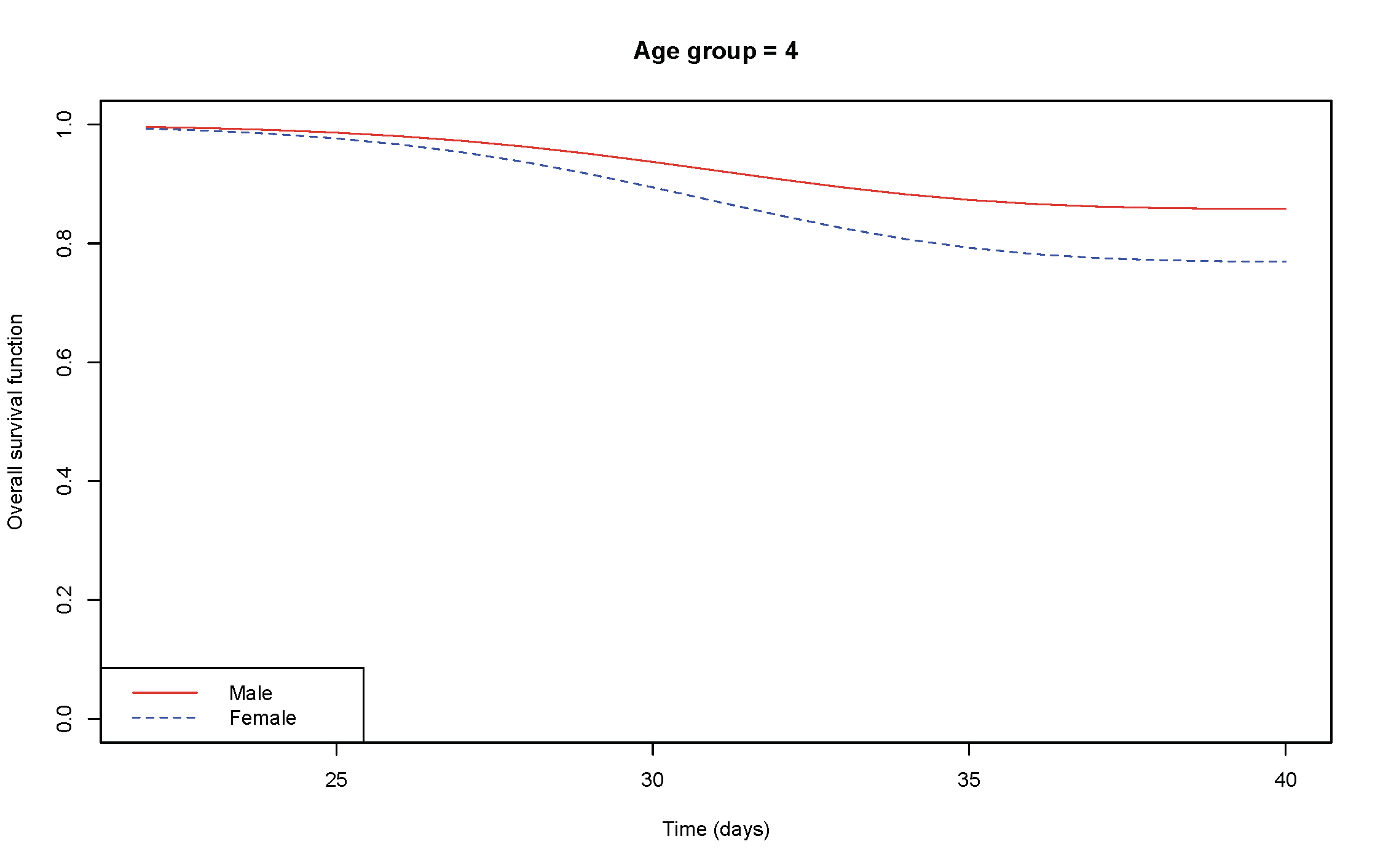}\\
		\includegraphics[scale=0.30]{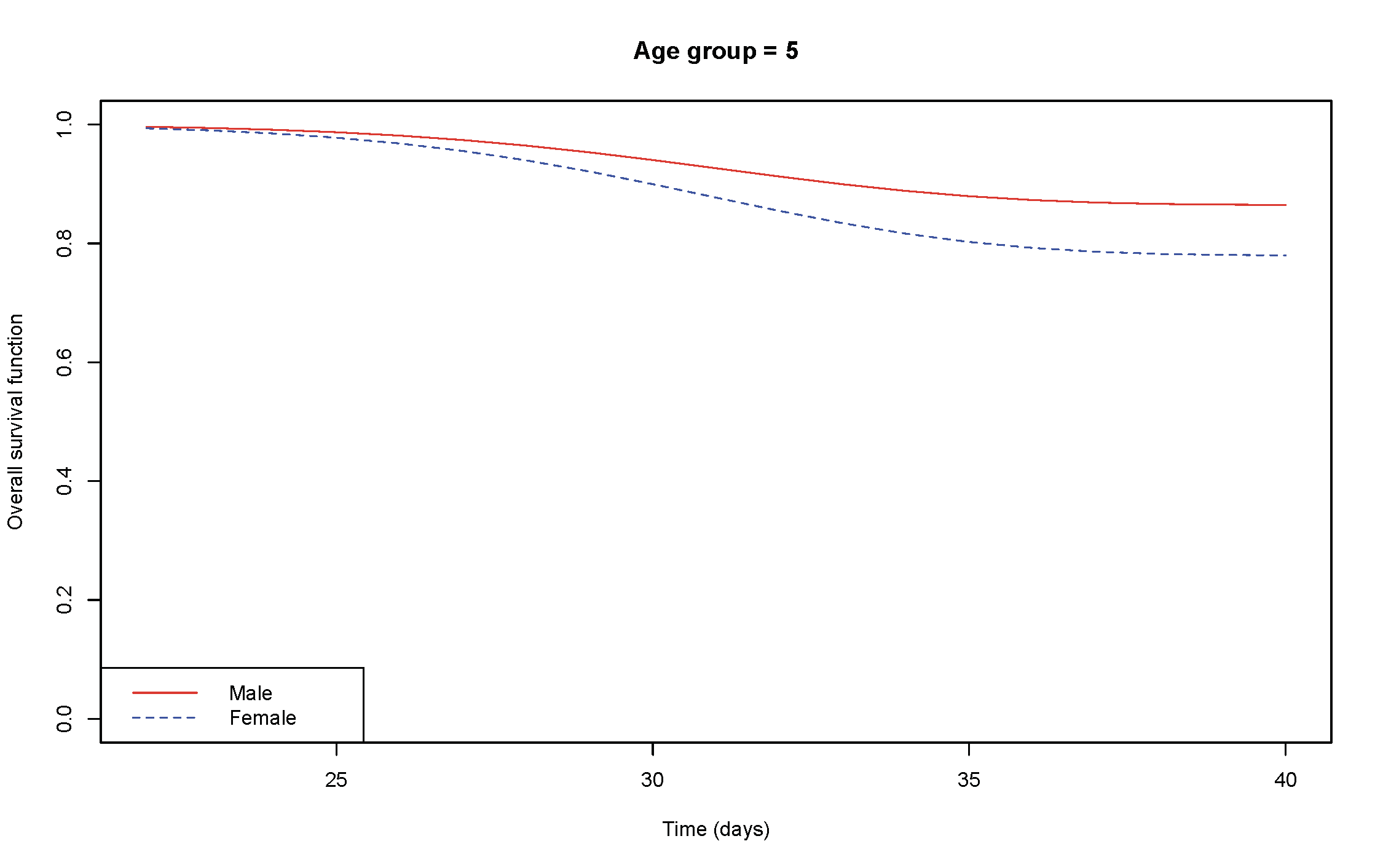}\\
\end{tabular}
  \caption{Predicted overall survival probability for different age groups and stratified by gender}
\label{figure:Sp}
\end{figure}

Figure \ref{figure:F2} presents a plot of the maximized log-likelihood function values against the selected values of the COM-Poisson shape parameter ($\nu$) for the real data. As we can clearly see, the values of the maximized log-likelihood function are very close to each other and almost indistinguishable.  Furthermore, Table \ref{table:real} shows some selected values of $\nu$ from the real data analysis along with the maximized log-likelihood, AIC, and BIC values. From Table \ref{table:real}, we can see there is very little difference between the models corresponding to different values of $\nu$. Therefore, we can say that any model within the COM-Poisson family is adequate for this data. For the remainder of this section, we will assume the data follows the geometric model since it has the highest log-likelihood value and  lowest AIC and BIC values. Under this assumption, Figure \ref{figure:Sp} presents the plots of estimated overall survival probabilities for different age groups and gender types. Note that for these plots we considered the number of exposures to be 6, which is the average number of exposures across all patients. Finally, in Table \ref{table:CP} we present the estimated cure probability for patients with different age groups and gender types, and considering the number of exposures to be 6. From both Figure \ref{figure:Sp} and Table \ref{table:CP} it is clear that for all age groups females tend to recover faster than males. 

\begin{center}
	\begin{table} [H]
		%\scriptsize
		\caption{AIC, BIC and maximized log-likelihood function $(\hat{l})$ values for different models}
		\begin{tabular*}{\textwidth}{@{\extracolsep{\fill}} l l l l}\\
			\hline
			Model & $\hat{l}$ & AIC & BIC \\                                 
			\hline
			COM-Poisson (geometric) & -85.7124 & 183.4247 & 198.7480 \\
			COM-Poisson $(\nu=0.5)$  & -85.7175 & 183.4350 & 198.7583\\
			COM-Poisson (Poisson)  & -85.7210 & 183.4420 & 198.7653\\
			COM-Poisson $(\nu=2)$  & -85.7252 & 183.4503 & 198.7736\\
			COM-Poisson (Bernoulli) & -85.7294 & 183.4588 & 198.7820\\ 
			\hline
		\end{tabular*}
		\label{table:real}
	\end{table}
\end{center}

\begin{table} [H]
	%\scriptsize
	\caption{Estimated cure probability for different age groups (AG) and gender types}
	\begin{tabular*}{\textwidth}{@{\extracolsep{\fill}} l l | l l | l l |  l l |  l l}\\ 
		\hline                                 
		 \multicolumn{2}{c}{AG=1} &  \multicolumn{2}{c}{AG=2} &  \multicolumn{2}{c}{AG=3} &  \multicolumn{2}{c}{AG=4} &  \multicolumn{2}{c}{AG=5} \\ \cline{1-10} 
		Male & Female & Male & Female & Male & Female & Male & Female & Male & Female \\
		\hline   
                 0.835 &0.734 & 0.843 &0.746 & 0.851 & 0.758 & 0.858 &0.769 & 0.865 & 0.780\\
                 \hline
	\end{tabular*}
	\label{table:CP}
\end{table}

\section{Conclusion and future work}

In this paper, we propose a new cure rate model for infectious diseases with discrete multiple exposures. Our model can capture both over-dispersion and under-dispersion that we usually encounter when modeling the count on pathogens at each exposure time. Instead of selecting any arbitrary distribution to model the number of pathogens at each exposure time, our model provides an approach to statistically test for the suitability of a chosen distribution within the wider family of COM-Poisson. In our simulation study, the model discrimination results clearly show the importance of selecting the correct distribution for the number of pathogens. Our proposed estimation procedure, based on right censored data, works very well in retrieving the true parameter values. The bias, RMSE and standard error are small and they further decrease with an increase in sample size. The coverage probabilities are also reasonably close to the true nominal level used. As a potential future work, we can look at the process of elimination of pathogens at each exposure time and develop the corresponding inference. It will also be of interest to consider the unobserved number of pathogens at each exposure time as frailty variables, as done in \citet{Tou08}, and use COM-Poisson as the frailty distributions. Another interesting future work is to assume a wider class of distributions for the pathogen promotion times, such as the generalized gamma distribution; see \citet{Pal20}. This will allow testing for the suitability of the commonly used distributions such as the Weibull, as considered in this paper. We are currently working on these and hope to report the findings in a future paper.

\section*{Acknowledgments}

The author expresses his thanks to two anonymous reviewers for providing useful comments and suggestions which led to this improved version of the manuscript.

\section*{Funding}

The research reported in this publication was supported by the National Institute Of General Medical Sciences of the National Institutes of Health under Award Number R15GM150091. The content is solely the responsibility of the author and does not necessarily represent the official views of the National Institutes of Health.

\section*{Conflict of Interest}

The author declares no potential conflict of interest. 

\section*{Data Availability Statement}

The real data is available on Kaggle. %The computational codes can be made available upon request.

\bibliographystyle{apacite}
\bibliography{Ref}

\section*{Appendix: Proof of Theorem 2.1}
	
	\begin{eqnarray*}
	S_{pop}(y) &=&  P[M_{t_0}=0,M_{t_1}=0]\\ 
	&&+ P[Z_{1,t_0}>y,...,Z_{M_{t_0},t_0}>y,M_{t_0}\geq 1,M_{t_1}=0]\\
	&&+ P[Z_{1,t_1}>y,...,Z_{M_{t_1},t_1}>y,M_{t_1}\geq 1,M_{t_0}=0]\\
	&&+ P[Z_{1,t_0}>y,...,Z_{M_{t_0},t_0}>y,Z_{1,t_1}>y,...,Z_{M_{t_1},t_1}>y,M_{t_0}\geq 1,M_{t_1}\geq 1]\\[1ex]
	&=&  P[M_{t_0}=0]P[M_{t_1}=0]\\ 
	&&+ P[Z_{1,t_0}>y,...,Z_{M_{t_0},t_0}>y,M_{t_0}\geq 1]P[M_{t_1}=0]\\
	&&+ P[Z_{1,t_1}>y,...,Z_{M_{t_1},t_1}>y,M_{t_1}\geq 1]P[M_{t_0}=0]\\
	&&+ P[Z_{1,t_0}>y,...,Z_{M_{t_0},t_0}>y,M_{t_0}\geq 1]P[Z_{1,t_1}>y,...,Z_{M_{t_1},t_1}>y,M_{t_1}\geq 1]\\[1ex]
	&=&P[M_{t_0}=0]P[M_{t_1}=0]\\ 
	&&+P[M_{t_1}=0]\sum_{k=1}^{\infty} P[Z_{1,t_0}>y,...,Z_{k,t_0}>y]P[M_{t_0}=k]\\
	&&+P[M_{t_0}=0]\sum_{j=1}^{\infty} P[Z_{1,t_1}>y,...,Z_{j,t_1}>y]P[M_{t_1}=j]\\
	&&+\sum_{k=1}^{\infty} P[Z_{1,t_0}>y,...,Z_{k,t_0}>y]P[M_{t_0}=k]\sum_{j=1}^{\infty}P[Z_{1,t_1}>y,...,Z_{j,t_1}>y]P[M_{t_1}=j]\\
	\end{eqnarray*}
	\begin{eqnarray*}
	&=&P[M_{t_0}=0]P[M_{t_1}=0]\\
	&&+P[M_{t_1}=0]\sum_{k=1}^{\infty}\lbrace S_{t_0}(y)\rbrace^kP[M_{t_0}=k]\\
	&&+P[M_{t_0}=0]\sum_{j=1}^{\infty}\lbrace S_{t_1}(y)\rbrace^jP[M_{t_1}=j]\\
	&&+\Bigg\lbrace\sum_{k=1}^{\infty}\lbrace S_{t_0}(y)\rbrace^kP[M_{t_0}=k]\Bigg\rbrace\Bigg\lbrace\sum_{j=1}^{\infty}\lbrace S_{t_1}(y)\rbrace^jP[M_{t_1}=j]\Bigg\rbrace.\\
	\end{eqnarray*}
	
	\noindent If we let 
	$$a_k=\lbrace S_{t_0}(y)\rbrace^kP[M_{t_0}=k]\hspace{1cm}, k=0,1,2,...$$
	and
	$$b_j=\lbrace S_{t_1}(y)\rbrace^jP[M_{t_1}=j]\hspace{1cm}, j=0,1,2,...,$$
	then, we know
	$$\sum_{j=0}^{\infty}\sum_{k=0}^{\infty}a_kb_j=a_0b_0+b_0\sum_{k=1}^{\infty}a_k+a_0\sum_{j=1}^{\infty}b_j+\sum_{j=1}^{\infty}\sum_{k=1}^{\infty}a_kb_j.$$
	Hence,
	\begin{eqnarray*}
	S_{pop}(y)&=& \sum_{j=0}^\infty\sum_{k=0}^\infty \{S_{t_0}(y)\}^kP[M_{t_0}=k] \{S_{t_1}(y)\}^jP[M_{t_1}=j] \\
	&=&\Bigg\lbrace\sum_{k=0}^{\infty}\lbrace S_{t_0}(y)\rbrace^kP[M_{t_0}=k]\Bigg\rbrace\Bigg\lbrace\sum_{j=0}^{\infty}\lbrace S_{t_1}(y)\rbrace^jP[M_{t_1}=j]\Bigg\rbrace\\
	&=&\Bigg\lbrace\sum_{k=0}^{\infty}\lbrace S_{t_0}(y)\rbrace^k\frac{1}{Z(\theta_{t_0},\nu)}\frac{\lbrace\theta_{t_0}\rbrace^k}{(k!)^v}\Bigg\rbrace\Bigg\lbrace\sum_{j=0}^{\infty}\lbrace S_{t_1}(y)\rbrace^j\frac{1}{Z(\theta_{t_1},\nu)}\frac{\lbrace\theta_{t_1}\rbrace^j}{(j!)^v}\Bigg\rbrace\\
	&=&\frac{1}{Z(\theta_{t_0},\nu)}\frac{1}{Z(\theta_{t_1},\nu)}\Bigg\lbrace\sum_{k=0}^{\infty}\frac{\lbrace \theta_{t_0}S_{t_0}(y)\rbrace^k}{(k!)^v}\Bigg\rbrace\Bigg\lbrace\sum_{j=0}^{\infty}\frac{\lbrace \theta_{t_1}S_{t_j}(y)\rbrace^j}{(j!)^v}\Bigg\rbrace\\
	&=&\frac{Z(\theta_{t_0}S_{t_0}(y),\nu)Z(\theta_{t_1}S_{t_1}(y),\nu)}{Z(\theta_{t_0},\nu)Z(\theta_{t_1},\nu)}\\
	\end{eqnarray*}
	as desired.

\end{document}